# Energy conversion of fully random thermal relaxation times

François Barriquand
proba5050@hotmail.com

**ABSTRACT.**
**Thermodynamic random processes in thermal systems are generally associated with one or several relaxation times, the inverse of which are formally homogeneous with energy. Here, we show in a precise way that the periodic modification of relaxation times during temperature-constant thermodynamic cycles can be thermodynamically beneficiary to the operator. This result holds as long as the operator who adjusts relaxation times does not attempt to control the randomness associated with relaxation times itself as a Maxwell 'demon' would do. Indirectly, our result also shows that thermal randomness appears satisfactorily described within a conventional quantum-statistical framework, and that the attempts advocated notably by Ilya Prigogine to go beyond a Hilbert space description of quantum statistics do not seem justified – at least according to the present state of our knowledge. Fundamental interpretation of randomness, either thermal or quantum mechanical, is briefly discussed.**

Since the pioneering works of Carnot and Clausius, thermal fluctuations have been described in rather negative terms, since, from a practical point of view, heat has been considered as a degenerate form of energy. One the positive side, however, thermodynamic fluctuations may also be considered as a unique tool allowing for the use of conventional thermal machines. To consider a very simple example, let us imagine a system under thermodynamic equilibrium possessing an eigenstate of lowest energy $E_1$ and corresponding occupation $p_1$. No Hamiltonian entirely controlled by the observer can compel the final occupation $p'_1$ of the lowest energy state to verify:

$$p'_1 \succ p_1 \qquad (1).$$

What 'pure' work alone cannot perform, fluctuations can achieve easily, however, through thermalisation at a lower temperature than T. When adjusted through time to interact with other thermal fluctuations, fluctuations may even be considered to be beneficial from the point of view of an external operator, as we now illustrate in two successive examples.

We first imagine a hollow ball filled with some diamagnetic liquid into which colloidal particles are kept in suspension. The colloidal particles are supposed to be paramagnetic, each particle possessing a single axis of highest paramagnetic susceptibility. We assume that when submitted to a magnetic field, the diamagnetic response of the liquid may be considered as nearly instantaneous, whereas the paramagnetic response of the colloidal particles is much slower, being conditioned by the Brownian motion of the liquid. We further suppose that for a given temperature T, when the ball remains at rest, the total static magnetic susceptibility of the ball is exactly zero. The ball is inserted at position x in a horizontal magnetic field B(x) oriented along xx'. The field B(x) possesses a magnetic gradient also oriented along xx'. Since the magnetic susceptibility of the ball is zero, the ball is not externally affected by the magnetic field. We now deposit the ball at the surface of a liquid where it can float and drift freely. The external magnetic field B(x) remains unaffected. The Brownian motion of the liquid outside the ball excites the rotational and translational degrees of freedom of the ball. Since the paramagnetic response of colloidal particles is slower than the diamagnetic response of the liquid inside the ball, the ball's rotation indirectly lowers its effective susceptibility. As a corollary result, in average, the ball drifts towards smaller fields values, i.e. from x towards x', and such movement can be converted into work at the benefit of the operator. If, after some work has been extracted from the system in that way, the operator lifts up the ball again and keeps it out of the liquid during a time long enough, all three rotational degrees of freedom of the ball being frozen, the effective magnetic susceptibility of the ball turns back to zero. From that moment, the operator can transfer the ball to higher x values at no energy cost. If, after this, the ball is again deposited on the liquid's surface, the ball starts to drift again towards smaller field values, etc. At first sight, it seems that some heat extracted from the liquid above which the ball is floating may be converted into work, while no external source of heat is used to heat or cool the whole system. Therefore, a careful examination of all energy transfers taking place during a cycle needs to be done. First of all, we may note that the external magnetic field needed for the experiment can be provided by a magnet that never moves, so that no work is directly transmitted from the magnet to the system or vice versa. What happens to the ball when it floats towards x', or when it is transported back towards x has already been described briefly. The most delicate steps correspond to the 'lifting' and the 'lowering' of the ball. Unfortunately, these steps are difficult to analyze accurately.

In order to remedy to this conceptual weakness, we need to find an analogous but more tractable experimental set-up, for which every step can be analyzed quantitatively. We now therefore propose to consider a kind of magnetic compass floating above a non-magnetic liquid whose susceptibility, either static or dynamic, is simply zero. The field of the compass is supposed to be horizontally oriented. For simplicity, we imagine that the compass is cylindrical in shape, its main axis of symmetry being oriented along the vertical axis zz'. No permanent magnetic field is imposed from outside on the system, and the earth's magnetic field is also supposed to be absent. The Brownian motion of the liquid induces the rotation of the horizontal magnetic field of the compass around the axis zz'. The typical time needed for the compass to make a half turn (randomly clockwise or anticlockwise) at temperature T can be noted as $T_{comp}(T)$. At time $t_0=0$, a box containing a gas whose only magnetic susceptibility we consider is due to the nuclear spins of its gaseous atoms is placed above the compass at a certain distance $d_0$. The reason we imagine that nuclear spins are the only source of magnetic susceptibility of the box is that at low fields, the relaxation time (usually noted $T_1$ in NMR) of such spins is known to be extremely long, easily exceeding days. Such $T_1$ can therefore be supposed to be so large that:

$$T_1 \succ\succ T_{comp} \qquad (2).$$

Under such regime, the compass turns too fast for the nuclear spins to follow the orientation of its magnetization efficiently. In average, the box containing the nuclear spins is less attracted by the compass than if the compass was fully at rest. Suppose now that during a time lasting from $t_0$ to $t_1$ ($t_1-t_0<<T_{comp}$), the moment of inertia of the compass is considerably augmented, no other parameter of the system being externally affected. This can be achieved if the compass possesses a built-in programmable machinery enabling it to move a massive (for simplicity, non-magnetic) part of its own structure away from its central vertical axis (above the surface of the liquid). Ideally, we may suppose that the moment of inertia of the compass becomes so large that the new typical time needed for the compass to make a half turn, noted $T'_{comp}$, verifies: $T'_{comp}>>T_1>>T_{comp}$. Starting from time $t_1$, the average thermal magnetization of the nuclear spins follows the random movement of the field of the compass quite efficiently. The box containing the nuclear spins becomes more attracted towards the compass. If, starting from a time $t_2$ (with $t_2-t_1>>T_1$), we allow the box to be externally moved down towards the compass, work can be extracted from the system during this time. The time when the box is finally stopped at some distance $d_1$ from the magnet ($d_1<d_0$) can be noted $t_3$. From $t_3$ to $t_4$ ($t_4-t_3=t_1-t_0$), the moment of inertia of the compass is reduced (using again the in-built machinery located inside the compass) by the same amount it was increased between $t_0$ and $t_1$. Starting then from $t_5$, with $t_5-t_4>>T_1$, the box is pulled upwards by the operator and reaches back the original position it had at time $t_0$. This demands some work from the operator, but less than the work extracted when the box was getting closer to the compass, since the box is less attracted by the compass from $t_4$ to $t_5$ than from $t_1$ to $t_2$. When the box, at a time $t_6$, finds back the original position it had at $t_0$, a new thermodynamic cycle can start again, etc. All the steps described in the cycle can be analysed in detail. This is true also for the two steps ($t_1-t_0$) and ($t_4-t_3$) during which the moment of inertia of the compass is modified. During these two steps, at first sight, the machinery built inside the compass hardly needs to spend any work. On closer analysis, random rotation of the compass must be considered: the higher the energy of rotation of the compass, the easier it is to enhance its moment of inertia (thanks to centrifugal force). By chance, the rotational energy of the compass happens to be statistically slightly higher at time $t_0$ (just before the moment of inertia is enhanced) than at $t_3$ (just before the moment of inertia is reduced), since the attractive potential of the nuclear spins, acting as a kind of 'brake' on the Brownian motion of the compass, is more effective at $t_3$ than at $t_0$. What happens when the moment of inertia of the compass changes is physically strongly analogous to what happened when the ball of our first example was lifted up or down. The advantage of our second example is simply that the energy balance of the system can be straightforwardly evaluated during these kinds of steps. There, thermal fluctuations appear generous enough to provide some work both to the observer manipulating the box of nuclear spins and to the machinery built inside the compass. In such an example, for the sake of clarity, we imposed that $T'_{comp} >> T_1 >> T_{comp}$. However, the mere condition $T'_{comp} > T_{comp}$ would suffice to authorize some energy conversion from heat to work, whatever the value of $T_1$. Quite similarly, in our first example, the crucial condition allowing for the conversion of heat into work during the translations of the ball along xx' was that the rotation of the ball in the liquid should be faster than its rotation above the liquid. Let us recall that in the presentation of our first example, we initially supposed that the rotation of the ball above the liquid was completely frozen, which amounts to say that the typical time associated with the rotation of the ball was infinite. Eventually, thermal fluctuations acting, not directly on fixed potentials controlled by the observer but rather on other thermal fluctuations, render possible the conversion of heat into work. The only necessary condition appears to be that the 'relaxation' time associated with one source of fluctuations may be varied at will by the operator. As our last example shows unambiguously, the operator does not necessarily need to spend any work in order to adjust such a thermodynamic time.

So far, no quantum statistics have been used in our analysis. But the question inevitably arises: since quantum statistics tell us that any eigenstate of energy E should be populated in proportion to:

$$\rho(E) \approx e^{-E/kT} \qquad (3)$$

when thermal equilibrium is attained, how can conversion of heat into work ever take place in that case? To find the correct explanation, we need to remember that thermal equilibrium, in fact, never corresponds to an exact thermalisation of 'pure' quantum states, but only to the statistical thermalisation of such states. Strictly speaking, thermalisation does not imply that any pure eigenstate of energy E is exactly populated according to eq.(3) at equilibrium, but that any wave function 'close enough' to a pure eigenstate of energy E should be populated in a proportion 'close enough' to the ideal ration provided by eq.(3) once equilibrium is reached. Even under thermal equilibrium, thermal fluctuations continually take place, so that the energy of any quantum state can only be known up to a precision of:

$$\delta E = \hbar / \text{Thermal relaxation time} \qquad (4).$$

Here, what we note as a *Thermal relaxation time* is the result of relevant random processes. In the case of nuclear spins, for instance, the statistical energy of their quantum states can only be defined up to a precision inversely proportional to $T_1$, where $T_1$ is the relaxation time of the nuclear spins. In our experimental schemes, we never try to 'order disorder' in a way similar to that considered by Maxwell[1], since such attempt would inevitably be accompanied by extra energy costs abundantly described in the literature. Rather, we accept randomness as it is, and only play with the relaxation times lying at the heart of random processes. It must be emphasized that entropy remains an authentic marker of the quantity of statistical disorder of our systems at any time. Our schemes would prove useless for recovering any thermally lost information. Only 'new' order may be introduced in the system.

Indirectly, our conclusions tend to indicate that, fundamentally, the flow of time which human beings perceive as irreversible is not ontologically connected with thermodynamic irreversibility. The attempts of Prigogine [2], who advocates the use of quantum wave functions outside of a Hilbert space in order to fix the arrow of time, seem difficult to reconcile with our results. Admittedly, the 'pure' eigenstates of a whole system (all external thermal or energy sources being included) always play experimentally a kind of ghostly role, since thermalisation, strictly speaking, cannot apply to them. Therefore, their very existence may ontologically be doubted, especially since no branch of experimental physics appears separable from thermodynamics. However, the hypothesis of their existence, which invites one to attribute apparent thermal irreversibility to the 'limited experimental capabilities' (some textbooks would rather say the 'limited knowledge', which seems less to the point) of an operator probing a macroscopic system, appears in good agreement with all experiments known so far including in principle the two ones considered above in this study.

The existence of 'pure eigenstates of the universe' and their significance for the interpretation of randomness nevertheless raises some conceptual difficulties which Prigogine has legitimately emphasised. Before concluding this article, we briefly evoke such difficulties. Our discussion successively comments epistemological, thermodynamic and quantum dynamical issues. All these issues are ultimately relevant in some way for thermodynamics.

On the epistemological level, let us consider the idea that 'pure eigenstates', or even possibly only one of them, could describe the whole universe, including therefore all human activities. Fundamentally, the epistemological problem raised by such kind of idea possesses a much older history than modern physics. Part of the problem was already raised by some Greek thinkers of the Antiquity, notably Parmenides and Heraclites, and reappeared later through the diverging ontological views of Plato and Aristotle. Also relevant to the problem is the point of view of the highly original third century B.C. Chinese thinker Gongsun Long [3] who says: 'The designation (of a thing) is not part of the world, whereas (designed) things are part of the world'. Basically, the problem at stake here lies in the coherence between matter and human language (and for us today, not only language, but also physics). Whatever the precise answer Gongsun Long effectively provided (his argumentation is actually very hard to follow in detail, and has received several contradictory interpretations), the problem he evoked remains relevant until today. Naively, we might be tempted to think today that some kind of totality should, by definition, provide us with a natural means to connect matter and language in some way. However, Russell's famous paradox [4] indicates that no *Set of all sets* can exist, since the *Set of all sets not containing themselves* would contain itself if and only if it didn't contain itself. Following the same line of reasoning, we might show that no *Theory encompassing all theories* can exist, and even that no *Thought of all thoughts* can exist. In this last example, Russell's paradox reaches a kind of climax since, as far as language is concerned, the *Thought of all thoughts* can always be considered as a thought. The *Thought of all thoughts* maybe meaningless or irrelevant, but in a way, by definition, it is a thought. Russell's paradox might be best understood within a dynamical perspective when realizing that any thought's assumed validity or meaning is always, humanly speaking, provisional, always offered to others for further confirmation, attestation, falsification, etc. Some thoughts, and among them some scientific theories, may prove *a posteriori* continuously valid. However, even in that case, such thoughts or theories may always be further tested, expanded, inserted within wider theoretical frames or linked together with other thoughts. Therefore, is the concept of 'pure eigenstate of the universe' acceptable in spite of its absolute, rather totalising character? Adopting this concept would amount to end our efforts directed at understanding the coherence between matter and language by concluding that the question is pointless, which at least temporarily seems hardly convincing. On the other hand, is epistemology itself *epistemologically legitimate* in discarding an experimentally satisfying physical theory? The wit of yet another ancient Chinese thinker, Zhuangzi, in the famous tale 'Three in the morning, four in the evening' [5] illustrates the sometimes limited scope of human language and, therefore, of epistemology if considered alone. In the end, from the epistemological point of view, it seems that the idea of 'pure eigenstates' of the whole universe, although quite disturbing, cannot be fully discarded *a priori*.

On the thermodynamic level, as has already been mentioned above, we see no clear reason to refuse the notion of 'pure eigenstates' of the universe. However, this opinion has only a temporary value, since quantum dynamics do not presently account for all physical phenomena in a unified way. Whether, for instance, any hypothetical future unification of gravity and quantum mechanics within a single theoretical frame will leave more space for chaotic randomness than quantum statistics alone presently provide is difficult to guess.

On the quantum dynamical level, the notion of 'pure eigenstates' of the universe becomes irrelevant when 'quantum collapse' events are allowed for. The phenomenon of quantum collapse remains physically poorly understood. An argument in favour of the natural occurrence of 'quantum collapse' events in nature lies in the fact that in a Young's slits experiments performed on particles detected one by one, quantum collapse never produces itself before an individual particle reaches the screen since otherwise no interference would be observed. If the conscience of an observer was responsible for quantum collapse, why should such conscience be unable to produce the collapse before particles reach the screen? Maybe, in fact, quantum collapse events are so common in nature that, for instance, all solar photons arriving on the earth experience a kind of 'quantum collapse' at their arrival. So far, however, nobody seems to know for sure if or with what kind of frequency quantum collapse events do happen in the universe. From the philosophical point of view, the tentatively ontologically indeterminist nature of 'quantum collapse' may have huge consequences. If 'quantum collapse' events really happen, present events may be considered to enjoy a certain degree of *autonomy* from the past of our universe. Such autonomy might provide a physical basis for the philosophical concepts of *novelty* and *contingency.* It may even help one to build up the concept of a *subject* (valid, at this level, for robots as well), which at the most basic level could be defined as the *legitimate historical reference of an action*. At a slightly more advanced level, subjects could even prove themselves capable to 'play dice with the universe'. Absence of ontological determinism and even concrete free will are not yet synonymous with meaningful freedom, however, since freedom, ontologically, might ideally require a certain degree of freedom from the self, which physically appears at first sight quite paradoxical. Further philosophical speculations remain outside the object of this study. From the point of view of physics, let us simply note that quantum collapse, which constitutes the only serious concept presently capable to challenge the relevance of the notion of 'pure eigenstates' of the universe, remains highly speculative in nature. What is more, nobody seems to know whether, or to what extent, quantum collapse events can influence the thermodynamics of the universe, as Von Neumann [6] early suggested they might do.

As a conclusion, we may simply note that the two examples provided above to illustrate how an operator may extract energy from some systems by varying the relaxation rate of thermal fluctuations acting on other fluctuations may pave the way for a completely new field of study and experiments. Meanwhile, entropy remains an authentic marker of the quantity of statistical disorder of macroscopic systems at any time. At best, only 'new' order may be introduced in a system, but lost information is lost anyway. Thermal irreversibility still exists, while 'novelty' may also be hoped for.